\documentclass[runningheads]{llncs}
\usepackage{graphicx}
\usepackage{xargs}                    
\usepackage[pdftex,dvipsnames]{xcolor}  
\usepackage[disable,colorinlistoftodos,prependcaption,textsize=tiny]{todonotes}
\newcommandx{\tddne}[2][1=]{\todo[inline,linecolor=green,backgroundcolor=green!25,bordercolor=green,#1]{#2}}
\newcommandx{\tdsec}[2][1=]{\todo[inline,linecolor=orange,backgroundcolor=orange!25,bordercolor=red,#1]{#2}}
\newcommandx{\unsure}[2][1=]{\todo[linecolor=red,backgroundcolor=red!25,bordercolor=red,#1]{#2}}
\newcommandx{\change}[2][1=]{\todo[linecolor=blue,backgroundcolor=blue!25,bordercolor=blue,#1]{#2}}
\newcommandx{\info}[2][1=]{\todo[linecolor=OliveGreen,backgroundcolor=OliveGreen!25,bordercolor=OliveGreen,#1]{#2}}
\newcommandx{\improvement}[2][1=]{\todo[linecolor=Plum,backgroundcolor=Plum!25,bordercolor=Plum,#1]{#2}}
\newcommandx{\thiswillnotshow}[2][1=]{\todo[disable,#1]{#2}}
\usepackage{hyperref}

\usepackage{blindtext}

\begin{document}
\title{Airway measurement by refinement of synthetic images improves mortality prediction in idiopathic pulmonary fibrosis}
\titlerunning{Airway measurement by synthetic images improves mortality prediction}
%
\author{Ashkan Pakzad\inst{1} \and
Mou-Cheng Xu\inst{1} \and
Wing Keung Cheung\inst{1}\and
Marie Vermant\inst{2,3}\and
Tinne Goos\inst{2,3}\and
Laurens J De Sadeleer\inst{2,3} \and
Stijn E Verleden\inst{2,4} \and
Wim A Wuyts\inst{2,3}\and
John R Hurst\inst{5} \and
Joseph Jacob\inst{1,5}}
\authorrunning{A. Pakzad et al.}
%
\institute{Centre for Medical Image Computing, University College London, UK \and
BREATHE, Department of Chronic Diseases and Metabolism, KU Leuven, Leuven, Belgium \and
Department of Respiratory Diseases, Unit for interstitial lung diseases, University Hospitals Leuven, Leuven, Belgium \and
Antwerp Surgical Training, Anatomy and Research Centre (ASTARC), Faculty of Medicine and Health Sciences, University of Antwerp, Antwerp, Belgium. \and
UCL Respiratory, University College London, UK \\
\email{a.pakzad@cs.ucl.ac.uk}\\
\url{https://ashkanpakzad.github.io}}
\maketitle              
\begin{abstract}
Several chronic lung diseases, like idiopathic pulmonary fibrosis (IPF) are characterised by abnormal dilatation of the airways. Quantification of airway features on computed tomography (CT) can help characterise disease progression.
Physics based airway measurement algorithms have been developed, but have met with limited success in part due to the sheer diversity of airway morphology seen in clinical practice. Supervised learning methods are also not feasible due to the high cost of obtaining precise airway annotations.
We propose synthesising airways by style transfer using perceptual losses to train our model, Airway Transfer Network (ATN). We compare our ATN model with a state-of-the-art GAN-based network (simGAN) using a) qualitative assessment; b) assessment of the ability of ATN and simGAN based CT airway metrics to predict mortality in a population of 113 patients with IPF.
ATN was shown to be quicker and easier to train than simGAN. ATN-based airway measurements were also found to be consistently stronger predictors of mortality than simGAN-derived airway metrics on IPF CTs. 
Airway synthesis by a transformation network that refines synthetic data using perceptual losses is a realistic alternative to GAN-based methods for clinical CT analyses of idiopathic pulmonary fibrosis.
Our source code can be found at \url{https://github.com/ashkanpakzad/ATN} that is compatible with the existing open-source airway analysis framework, AirQuant.

\end{abstract}
\section{Introduction} \label{sec:intro}

Chronic lung disease is one of the leading causes of morbidity and mortality across the world. As smoking rates in the developing world increase, the prevalence of chronic lung disease is set to rise. Interstitial lung diseases (ILD) are characterised by inflammation and scarring of the lung and the incidence of ILD continues to increase \cite{xie_trends_2020}. 

A subset of ILDs are characterised by lung fibrosis, with idiopathic pulmonary fibrosis (IPF) having the worst prognosis of all the fibrosing ILDs \cite{flaherty_idiopathic_2006}. In IPF the airways are pulled open by fibrotic contraction of the surrounding connective tissue. Computed tomography (CT) imaging is used to visualise airway structure. In IPF the presence of dilated airways in the lung periphery on CT, termed traction bronchiectasis, is a disease hallmark. 

When assessing disease severity in IPF, physiologic measurements are typically used. However these are associated with a degree of measurement variability. It has been postulated that combining imaging measures of airway abnormality with lung function measurements could help improve estimation of disease severity in IPF \cite{pakzad_radiology_2022}. Importantly, better measures of disease severity would benefit cohort enrichment of subjects into therapeutic trials. 

Lung damage in IPF progresses from the distal lung towards the centre of the lung \cite{lederer_idiopathic_2018}. As a result, the earliest signs of lung damage are seen in the smaller airways. Yet these airways are typically the most challenging to quantify. Airway measurement is complicated by partial volume effects that result in smaller airways having a blurred contour to their walls. Measurement challenges are compounded by variations in CT image acquisition including different reconstruction kernels, scan parameters and scanner models as well as the underlying pathology affecting the lung. 

Physics based airway measurement algorithms tend to perform sub optimally when measuring the lumens of small airways \cite{kiraly_virtual_2005,estepar_accurate_2006}. Identifying airway walls can also be challenging. Airway paths often run in tandem with those of the pulmonary artery. Consequently, in regions when the pulmonary artery abuts the airway wall, identification of the contour of the outer airway wall is compromised.

\subsection{Related work} \label{sec:related}
Deep learning frameworks have been applied to the measurement of airways in the lung in a bid to improve measurement accuracy. However, these machine learning methods are extremely data hungry and can be biased towards the training data sample \cite{hofmanninger_automatic_2020}. Synthetic data by way of generative models has been employed to improve the training of deep learning models. This helps overcome the data limitations that are ubiquitous to medical imaging studies \cite{willemink_preparing_2020}.

A state of the art method in measuring airway lumen radius and wall thickness on CT imaging, simGAN \cite{nardelli_generative-based_2020,shrivastava_learning_2017}, takes labelled simplistic representations of airway patches (synthetic images) and aims to transforms them in to the emulations of real airways by generative adversarial training (GAN) \cite{goodfellow_generative_2014}. These are then used for supervised training of a convolutional neural regressor (CNR) which learns to measure airway radius and wall thickness and ultimately to run inference on real CT images.

The driving loss for realism in simGAN is cross-entropy loss computed on the classifications of the discriminator. For successful synthetic refinement by image transformation, the synthetic and refined images must have good correspondence in their shared label. To this end, a per-pixel $\|l\|_{1}$ regularisation loss is applied between input and output of the refiner. 

GAN training is inherently unstable with mode collapse complicating and lengthening training times. As an alternative strategy, in this paper we propose the first use of perceptual losses to generate labelled synthetic airway images. Perceptual loss functions have been applied to image style transfer and super-resolution tasks\cite{johnson_perceptual_2016}. We explore the clinical benefits of learning from perceptual loss generated synthetic data in mortality prediction.

\section{Methods}

In the first part of our study we generate synthetic airway patches that demonstrate realistic airway characteristics. In tandem, we segment the airways on clinical CT scans of a cohort of IPF patients. We train our Airway Transfer Network (ATN) to transform our synthetic images to refined images across our synthetic and real datasets by optimising for perceptual losses. We then compare the results of ATN with simGAN. A CNR is trained on the resultant refined datasets for the purpose of inference on real CT airways. 

We compare the two refiner models qualitatively. We compare ATN and simGAN against the full width at half maximum edgecued segmentation limited (FWHMesl) technique as implemented in \cite{quan_tapering_2018}, originally by \cite{kiraly_virtual_2005}. The FWHMesl technique is widely used in the literature as the reference for comparison of previous airway measurement methods \cite{nardelli_generative-based_2020,xu_hybrid_2015,gu_computerized_2013}. In our clinical comparison, we examine which of the three methods of airway measurement provides the best and most consistent association with mortality on CT scans of patients with IPF.

Airway segmentation was performed using a 2D dilated U-Net \cite{yu_multi-scale_2016} trained on CT scans in 25 IPF and healthy individuals \cite{pakzad_evaluation_2021}. We extract orthogonal airway patches for all segmented airways. We parameterise airway labels as two ellipses that share centre and rotation, resulting in 7 parameters for each patch: inner airway wall major and minor axis radii $R_{A}$ and $R_{B}$; outer airway wall major and minor axis radii $W_{A}$ and $W_{B}$; centre coordinates {$C_{x}$} and $C_{y}$; and rotation $\theta$. Due to the phase in $\theta$, for the purposes of CNR training the rotation angle is converted into a double angle representation \cite{kluvanec_using_2022}. 

Once the refiner model has been trained, its output is used to train a CNR by supervised learning to regress to target airway labels. The inner and outer airway wall measures are then derived. All deep learning methods were implemented in pytorch \cite{paszke_pytorch_2019} and CT image processing was done using the open source airway analysis framework known as AirQuant \cite{pakzad_evaluation_2021}. We release our code open source\footnote{\url{https://github.com/ashkanpakzad/ATN}}.

\subsection{Airway Synthesis} \label{sec:synthesis} 
Details of airway parameters and synthesis pipeline have been previously described \cite{nardelli_generative-based_2020}. Airway characteristics are sampled from a set of distribution parameters informed by \cite{weibel_morphometry_1965}. We deviate from these parameters in two ways. First, we use an airway lumen radius (LR) interval of [0.3, 6] to permit measurement of smaller airways. Second, we use an airway wall thickness [$0.1 \cdot LR + 0.2$,$0.3\cdot LR+0.8$] mm to reflect the lack of airway wall thickening in IPF. We add four further parameters: (i) parameters for the airway centre determined by a normal distribution $X\sim N(0,1)$ mm to account for airway skeletons that are not perfectly positioned within the centre of the airway lumen. (ii) $p=0.4$ that an adjacent airway of similar diameter is randomly added. This is performed to accommodate airway patches close to airway bifurcations and to train the CNR to correctly identify the airway in the centre of the patch. (iii) We model our airways as ellipsoids, we achieve this by an ellipsoidness characteristic, sampled from a uniform distribution, $X\sim U(0.9,1)$ which determines the ratio in major and minor radii of the ellipse. (iv) Uniformly random rotation applied to the airway in the horizontal axis. We include our synthetic airway generator and configuration parameters in the open-source code repository.

\subsection{Perceptual Losses}
We implement perceptual losses for computing high level perceptual differences between synthetic and real images as described by \cite{johnson_perceptual_2016}. These losses are computed by comparing the activations in particular layers, $j$ of a pretrained convolutional neural network (CNN), $\phi$ between a pair of images. Different activation layers of a trained CNN learn to represent different image features on the same sampled patch. In minimising for perceptual losses we are looking to reduce the differences in the activation of these layers between the refiner output and some objective image. For each calculation of perceptual losses on a synthetic input image, $x$ we have a refiner prediction, $\hat{y}$. As a modification of the original style transfer implementation \cite{johnson_perceptual_2016}, a randomly chosen real image is selected as the style target, $y_{s}$. Perceptual losses are then calculated and summed for different layers $\phi_{j}$. 

We utilise feature reconstruction loss. This is defined as the mean euclidean distance between activations of the input and output images of the refiner, where $C$, $H$, and $W$ are the number of channels, height and width of layer $j$ respectively. We use a VGG-16 \cite{simonyan_very_2015} network pretrained on the ImageNet dataset \cite{deng_imagenet_2009} in our calculations of style and feature losses.

\begin{equation}
    l^{\phi,j}_{feat}(\hat{y},x) = \frac{1}{C_{j}H_{j}W_{j}} \|\phi_{j}(\hat{y}) - \phi_{j}(x)\|_{1}
\end{equation}

We also employ style reconstruction loss, which considers those features that tend to be activated together between the refiner output and the given style target image, a random real airway, where $G^{\phi}_{j}$ is the gram matrix for a given layer $j$ of $\phi$ as described in \cite{gatys_neural_2015}.  

\begin{equation}
    l^{\phi,j}_{style}(\hat{y},y_{s}) = \frac{1}{C_{j}H_{j}W_{j}} \|G^{\phi}_{j}(\hat{y}) - G^{\phi}_{j}(y_{s})\|_{1}
\end{equation}

\subsection{Clinical Data} \label{sec:clin}

We examined CT images from 113 IPF patients diagnosed at the University Hospitals Leuven, Belgium. CTs were evaluated by an experienced chest radiologist (author JJ) for quality i.e. absence of breathing artefacts and infection. The quality of the automated segmentation was also visually inspected to ensure contiguous airway segmentations without oversegmentation blowouts. Airway segmentations were also required to reach the sixth airway generation in the upper and lower lobes to be selected for analysis. Pulmonary function tests were considered if they occurred within 90 days of the CT scan: Forced Vital Capacity (FVC, n=111)); diffusing capacity of the lung for carbon monoxide (DLco, n=103).  

The trachea and first generation bronchi were excluded from analysis. We define an airway segment as the length of airway that runs between airway branching points or an airway endpoint. All airway segments were pruned by 1 mm at either end to avoid bifurcating patches. $80\times80$ pixel size orthogonal airway patches were linearly interpolated with a pixel size of $0.5\times0.5$ mm from the CT at 0.5 mm intervals along each segment. This resulted in a final set of 546,790 real CT-derived airway patches. A synthetic dataset of 375,000 patches was generated to train our refiner.


27\% of patients were female. 74\% of patients had smoked previously. The median patient age was 71, with 57\% of patients having died. All patients had received antifibrotic drug treatment. 

Measures of intertapering, intratapering \cite{kuo_airway_2020} and absolute airway volume were derived from the airway measurements for each airway segment. \textbf{Segmental intertapering} represents the relative difference in diameter of an airway segment when compared to its parent segment.
Segmental intertapering is calculated as the difference in mean diameter, $\bar{d}$ of an airway segment and its parent segment, $\bar{d_{p}}$, divided by the mean diameter of the parent segment. \textbf{Segmental intratapering} is the gradient of change in diameter of the airway segment relative to the diameter of the origin of the segment\footnote{Segments are considered to be oriented from the centre of the lung to the periphery. Accordingly, measurement of the airway origin beings at the end closest to the trachea}. Segmental intratapering is computed by dividing the gradient, $m$ by the zero-intercept, $c$ of a line $y=mx+c$ fitted to the diameter measurements of an airway segment. \textbf{Segmental volume} is computed by summing area measurements along an airway segment, and multiplying this value by the measurement interval, i.e. an integration of area along the segment's length.

\begin{equation}
    intertapering = \frac{\bar{d_{p}}-\bar{d}}{\bar{d_{p}}} 
\end{equation}

\begin{equation}
    intratapering = \frac{-m}{c} 
\end{equation}

Univariable and multivariable Cox proportional hazards models were used to examine patient survival. Multivariable models included patient age (years), gender, smoking status (never vs ever) and either FVC or DLco (as measures of disease severity) as covariates. The goodness of fit of the model was denoted by the concordance index \cite{harrell_multivariable_1996}. A p-value of $<$0.05 was considered statistically significant. 

\subsection{Implementation details} \label{sec:implementation}

We use the same refiner architecture as in \cite{shrivastava_learning_2017,nardelli_generative-based_2020}, the refiner is a purely convolutional network with four repeating 3x3, 64 feature ResNet blocks \cite{he_deep_2015}. The measurement CNR, described in \cite{nardelli_generative-based_2020}, is a convolutional network that feeds into two fully connected layers to learn the airway ellipse parameters.
Instead of the custom CNR loss described in \cite{nardelli_generative-based_2020}, we implemented a mean square error (MSE) loss for regressing to the airway ellipse parameters.

Synthetic images were generated to $0.5\times0.5$ mm pixel size making $80\times80$ pixel patches, corresponding to the real patch generation noted in section \ref{sec:clin}. All images were standardised and augmented on the fly, adding random Gaussian noise [25,25] Hounsfield units, random levels of Gaussian blurring with standard deviation scalled in the interval [0.5, 0.875] and random flipping ($p=0.2$). We apply random scaling on real images only, in the interval [0.75,1.25] to increase diversity in airway size. Finally, a centre crop was applied to make a $32\times32$ pixel input patch. 

Both simGAN and ATN models were trained for 10000 steps, where the simGAN refiner had 50 training iterations and the discriminator 1 iteration for every 1 step. The simGAN discriminator was implemented as described in the original method, with a memory buffer and local patch discrimination \cite{shrivastava_learning_2017}. 
\section{Results}
\begin{figure}[!ht]
    \centering
    \includegraphics[width=1\textwidth]{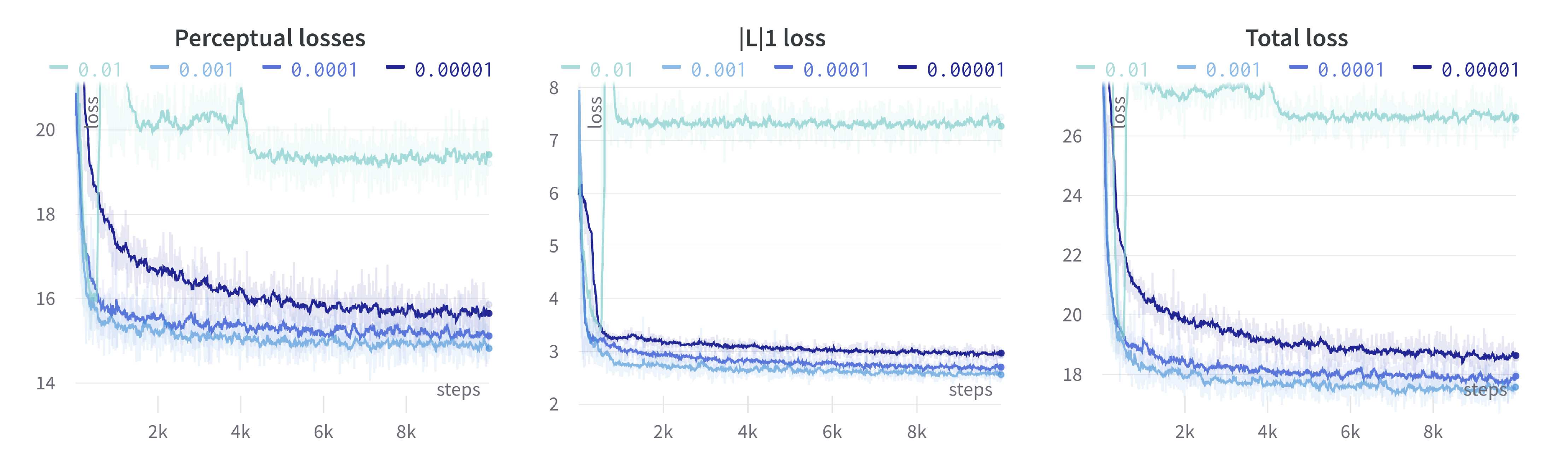}
    \caption{Comparison of loss minimisation for training the airway transformation network with different learning rates, demonstrating best model convergence at 0.001.}
    \label{fig:lr}
\end{figure}

We implemented all training on an NVIDIA GeForce RTX 2070 graphical processing unit with a batch size of 256, learning rate of 0.001, $\|l\|_{1}$ regularisation factor in range of [0.0001, 0.1]. Figure \ref{fig:lr} shows training convergence for different learning rates where simGAN and ATN took 14 and 0.6 hours respectively to train. We qualitatively found that both simGAN and ATN produced refined images of optimal quality with a $\|l\|_{1}$ regularisation factor of 0.01. 

Style-transfer from paintings to natural images show that larger-scale structure is transferred from the target image when training on losses of higher layers \cite{johnson_perceptual_2016}. In order to maintain label correspondence between refiner input and output, we similarly only use the feature loss using the \texttt{relu3\_3} activation layer. Style loss is computed from the two lower \texttt{relu1\_2}, \texttt{relu2\_2} activation layers only \footnote{higher activation layers are considered in the supplementary material}. Figure \ref{fig:refined} demonstrates qualitative results of our airway refinement method.

\begin{figure}[!ht]
    \centering
    \includegraphics[width=1\textwidth]{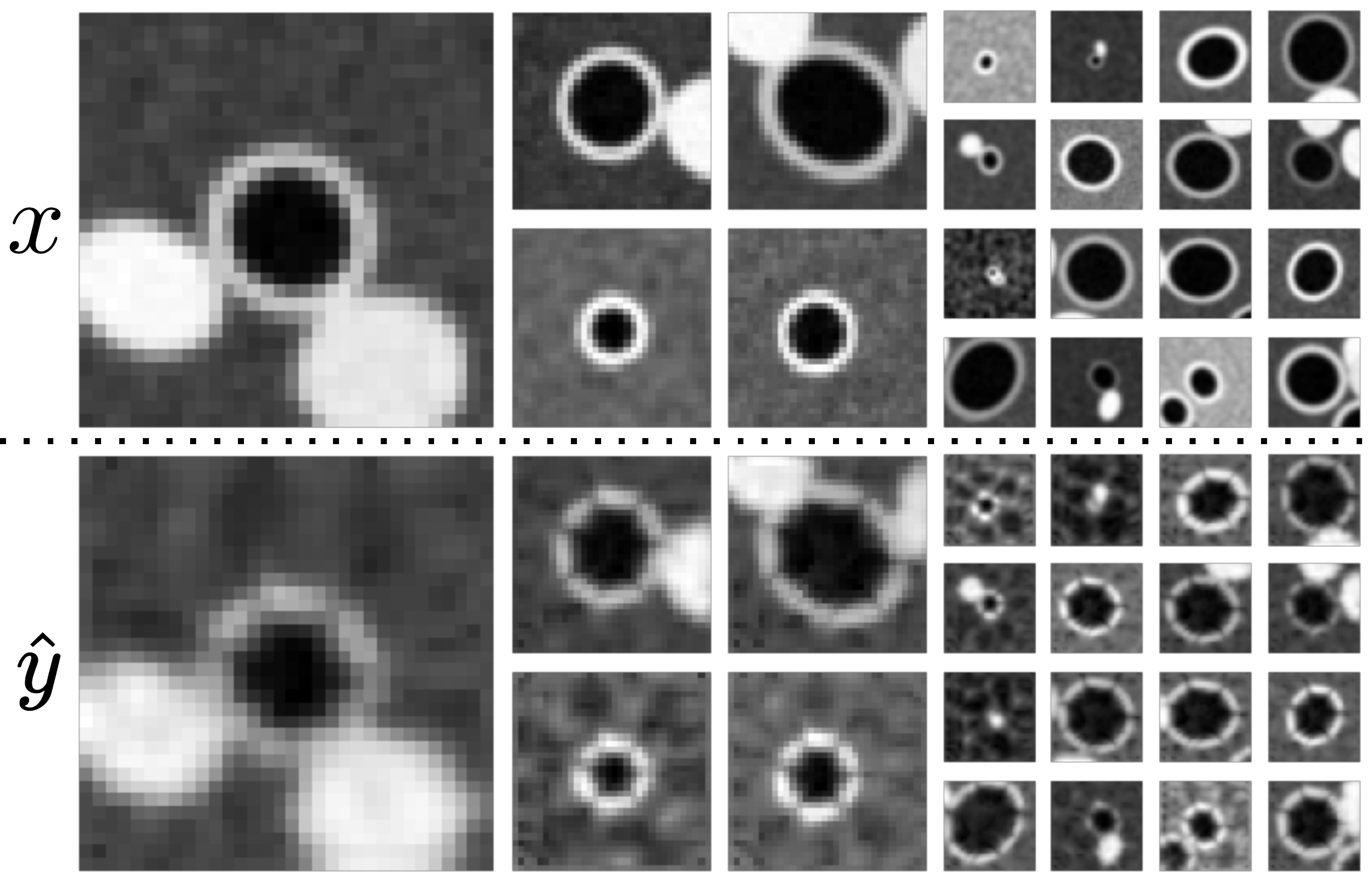}
    \caption{Uncurated set of synthetic images $x$ and output $\hat{y}$ of our airway transformation network in the same relative position below. Our model was trained to minimise perceptual losses. Airways are all represented at different scales.}
    \label{fig:refined}
\end{figure}

\begin{figure}[!ht]
    \centering
    \includegraphics[width=1\textwidth]{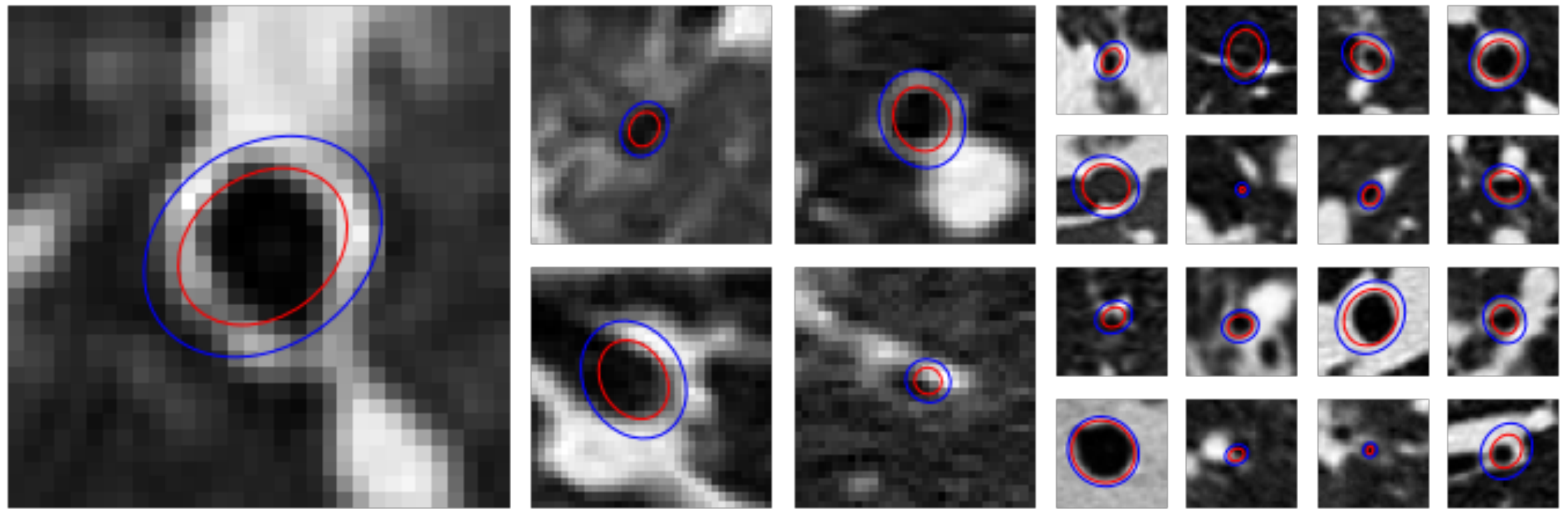}
    \caption{Uncurated inference on real airway patches performed by our airway measurement regressor network. The network was trained on refined synthetic data from our proposed airway transformation network, which minimises perceptual losses. The inner red ellipse delineates the inner airway wall and the outer blue ellipse, the outer airway wall. Airways are all presented at different scales.}
    \label{fig:inference}
\end{figure}

The CNR was trained with batch size in the interval [256,2000] and learning rate of 0.001. Batch size of 2000 was chosen for its speed, and converged at around 40 epochs within one hour. The CNR achieves comparable results on ATN and simGAN refined images. Figure \ref{fig:inference} demonstrates qualitative results of our ATN method on real CT data. Table \ref{tab:clinical} shows results of the Cox regression survival analyses.

\begin{table}[!ht]
\centering
\caption{Cox proportional hazards results comparing mortality prediction of airway biomarkers derived by different measurement methods.}
\label{tab:clinical}
\begin{tabular}{|lcccccc|}
\hline
\multicolumn{1}{|l|}{}                                            & \multicolumn{2}{c|}{Univariable (n=113)}              & \multicolumn{2}{c|}{DLCo  (n=103)}                    & \multicolumn{2}{c|}{FVC  (n=111)}   \\
\multicolumn{1}{|l|}{Method}                                      & C index       & \multicolumn{1}{c|}{p-value}          & C index       & \multicolumn{1}{c|}{p-value}          & C index       & p-value             \\ \hline
\multicolumn{7}{|c|}{\textbf{Volume}}                                                                                                                                                                                   \\ \hline
\multicolumn{1}{|l|}{FWHMesl\cite{quan_tapering_2018}}            & 0.61          & \multicolumn{1}{c|}{0.00190}          & 0.67          & \multicolumn{1}{c|}{0.03031}          & 0.68          & 0.03965             \\
\multicolumn{1}{|l|}{simGAN\cite{nardelli_generative-based_2020}} & 0.65          & \multicolumn{1}{c|}{0.00006}          & 0.68          & \multicolumn{1}{c|}{0.00233}          & 0.70          & 0.00086             \\
\multicolumn{1}{|l|}{ATN(ours)}                                   & \textbf{0.67} & \multicolumn{1}{c|}{\textbf{0.00001}} & \textbf{0.69} & \multicolumn{1}{c|}{\textbf{0.00013}} & \textbf{0.71} & \textbf{$<$0.00001} \\ \hline
\multicolumn{7}{|c|}{\textbf{Intertapering}}                                                                                                                                                                            \\ \hline
\multicolumn{1}{|l|}{FWHMesl\cite{quan_tapering_2018}}            & 0.55          & \multicolumn{1}{c|}{0.07009}          & 0.66          & \multicolumn{1}{c|}{0.14999}          & 0.68          & 0.08744             \\
\multicolumn{1}{|l|}{simGAN\cite{nardelli_generative-based_2020}} & 0.60          & \multicolumn{1}{c|}{0.00925}          & 0.67          & \multicolumn{1}{c|}{0.03460}          & 0.69          & 0.04764             \\
\multicolumn{1}{|l|}{ATN(ours)}                                   & \textbf{0.62} & \multicolumn{1}{c|}{\textbf{0.00084}} & \textbf{0.69} & \multicolumn{1}{c|}{\textbf{0.00062}} & \textbf{0.70} & \textbf{0.00052}    \\ \hline
\multicolumn{7}{|c|}{\textbf{Intratapering}}                                                                                                                                                                            \\ \hline
\multicolumn{1}{|l|}{FWHMesl\cite{quan_tapering_2018}}            & 0.55          & \multicolumn{1}{c|}{0.33623}          & 0.66          & \multicolumn{1}{c|}{0.93103}          & 0.69          & 0.63837             \\
\multicolumn{1}{|l|}{simGAN\cite{nardelli_generative-based_2020}} & 0.59          & \multicolumn{1}{c|}{0.09232}          & 0.67          & \multicolumn{1}{c|}{0.35460}          & 0.69          & 0.48513             \\
\multicolumn{1}{|l|}{ATN(ours)}                                   & \textbf{0.63} & \multicolumn{1}{c|}{\textbf{0.00026}} & \textbf{0.68} & \multicolumn{1}{c|}{\textbf{0.00208}} & 0.69          & \textbf{0.00192}    \\ \hline
\end{tabular}
\end{table}

\section{Conclusion}

We present a learning based airway measurement method trained on a transformation network that refines synthetic data using perceptual losses. Our model ATN was compared with a state-of-the-art model simGAN \cite{nardelli_generative-based_2020} and a physics based method FWHMesl. When assessing the clinical utility of ATN, we found that it was the strongest predictor of survival across all three airway biomarkers. We found that our method trains faster and with minimal complications, unlike a GAN framework. We expect future work to consider the versatility of such a method, for example examining airways in patients with different pathologies, different scanner parameters and potentially on higher scale imaging such as micro-CT studies of the lungs.

\section*{Acknowledgements}
This research was funded in whole or in part by the Wellcome Trust [209553/Z/17/Z]. For the purpose of open access, the author has applied a CC-BY public copyright licence to any author accepted manuscript version arising from this submission. AP is funded jointly by the Cystic Fibrosis Trust and EPSRC i4health, centre for doctoral training studentship. JJ was supported by a Wellcome Trust Clinical Research Career Development Fellowship and the NIHR UCLH Biomedical Research Centre, UK.

\bibliographystyle{splncs04}
\bibliography{ATNpaper}

\appendix
\section{Supplementary Material}

\subsection{Style loss ablation study}

\begin{figure}[!ht]
    \centering
    \includegraphics[width=1\textwidth]{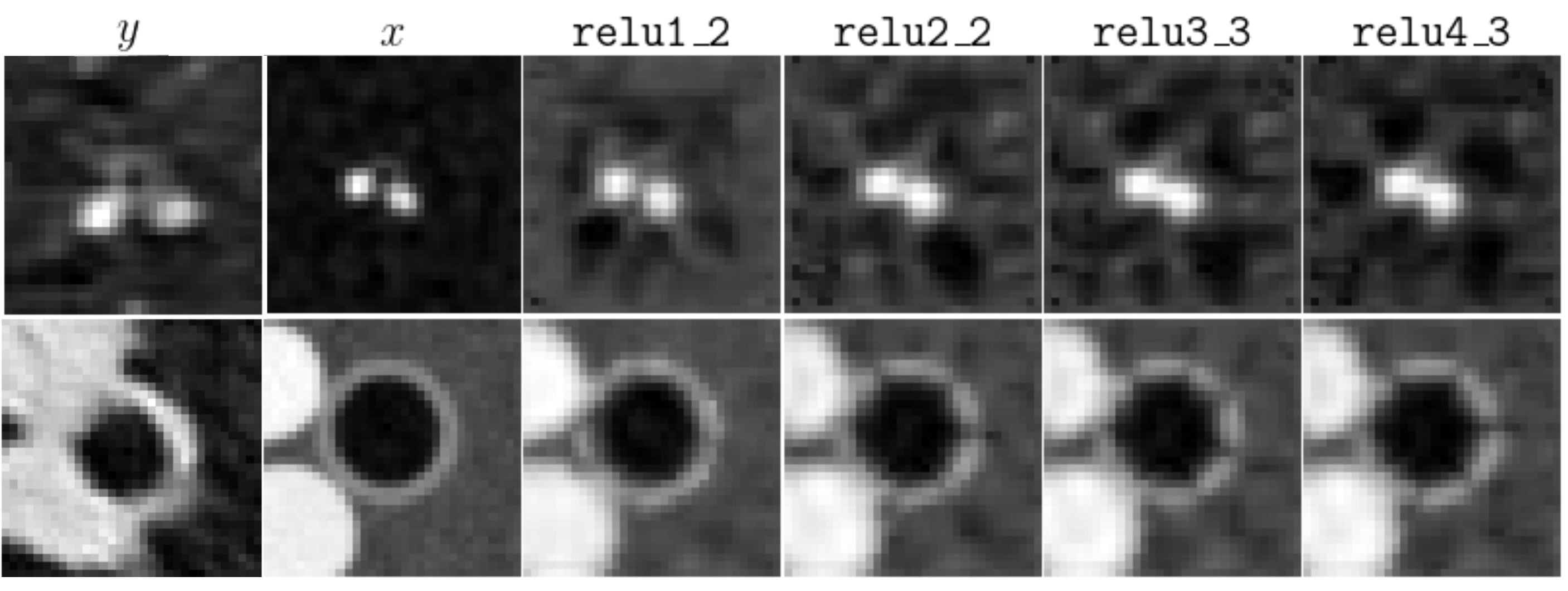}
    \caption{Similar to \cite{johnson_perceptual_2016}, for two example synthetic inputs $x$ we demonstrate the effect of minimising for style reconstruction loss using different activation layers of the pretrained VGG-16 loss network. Losses are accumulated with each higher layer. $y$ taken from a real clinical CT scan is similar in appearance to the synthetic images.}
    \label{fig:style}
\end{figure}

We consider the effects of minimising style loss with higher layers of $\phi$ in Figure \ref{fig:style}. The pretrained VGG-16 layers of imagenet from low to high was \texttt{relu1\_2}, \texttt{relu2\_2}, \texttt{relu3\_3} and \texttt{relu4\_3}. The style reconstruction losses are summed consecutively for each higher layer considered. This results in the final loss computed for \texttt{relu4\_3} being the previous 3 activation layers' losses.

We find the most significant qualitative difference occurs when moving from \texttt{relu1\_2} to \texttt{relu2\_2}, where \texttt{relu1\_2} appears quite smooth. The addition of each higher layer appears to add higher frequency details. The final layers introduce visually discernible larger-scale spatial changes, particularly in the smaller airways. Informed by this experiment and previous style transfer experiments by \cite{johnson_perceptual_2016}, we chose to only train on style reconstruction losses computed from \texttt{relu1\_2} and \texttt{relu2\_2}.

\end{document}